# Performance assessment of CsI(Tl) screens on various substrates for X-ray imaging[*]

FENG Zhao-Dong(冯召东)[1,3,4] JIANG Peng(蒋鹏)[2] ZHANG Hong-Kai(张红凯)[1,3,4] ZHAO Bo-Zhen(赵博震)[1,3] QIN Xiu-Bo(秦秀波)[1,3;1)] WEI Cun-Feng(魏存峰)[1,3] LIU Yu(刘宇)[1,3] WEI Long(魏龙)[1,3]

1 Key Laboratory of Nuclear Radiation and Nuclear Energy Technology, Institute of High Energy Physics, Chinese Academy of Sciences, Beijing 100049, China

2 The College of Nuclear Technology and Automation Engineering, Chengdu University of Technology, Chengdu 610059, China

3 Beijing Engineering Research Center of Radiographic Techniques and Equipment, Beijing 100049, China

4 University of Chinese Academy of Sciences, Beijing 100049, China

**Abstract:** Thallium-doped cesium iodide (CsI(Tl)) screens are widely used in X-ray imaging devices because of the columnar structure of CsI(Tl) layer, but few reports focus on the optical role of the substrate in the screen system. In this paper, four substrates including fused silica ($SiO_2$), silver-film coated $SiO_2$, graphite (C) and fiber optic plate (FOP) are used to fabricate CsI(Tl) screens by thermal evaporation. Their imaging performance is evaluated by relative light output (RLO), modulation transfer function (MTF), normalized noise power spectrum (NNPS) and noise equivalent quanta (NEQ). The results reveal that although CsI(Tl) film on graphite plate yields images with the lowest light output, it presents relatively higher spatial resolution and better signal-to-noise characteristics. However, films on $SiO_2$ plate obtain low MTF but high NNPS curves, whether or not coated with silver film. Furthermore, scintillation screens on FOP have bright images with low NNPS and high NEQ, but have the lowest MTF. By controlling the substrate optical features, CsI(Tl) films can be tailed to suit a given application.

## 1. Introduction

In the last few decades, thallium-doped cesium iodide (CsI(Tl)) scintillation screens have had widespread application in X-ray imaging devices for X-ray microscopy [1], digital radiography [2] and computed tomography (CT) [3], because their micro-columnar structure can depress the lateral scattering of scintillation photons. However, the compromise between X-ray absorption efficiency and spatial resolution restricts the applications of CsI(Tl) screens. It has been demonstrated that the imaging performance is affected by film evaporation parameters including substrate temperature, deposition rate, chamber pressure and Tl doping concentration. Cha et al. [4] deposited CsI(Tl) scintillation films on glass plate for different Tl concentrations, and evaluated their imaging performance. Fedorov et al. [5] fabricated CsI(Tl) layers on LiF and glass substrates, and examined their scintillation efficiency, crystal structure and spatial resolution. Shinde et al. [6] deposited CsI(Tl) films on silicon substrate in the thickness range between 10 nm and 3 μm, and focused on the temperature dependent

*Supported by National Key Scientific Instrument and Equipment Development Project (2011YQ03011205, 2013YQ03062902) and Key Program of the National Natural Science Foundation of China (U1332202)

1） E-mail: qinxb@ihep.ac.cn

photoluminescence of CsI(Tl) films. Yao et al. [7] fabricated CsI(Tl) films on glass substrate covered by a pre-deposited CsI layer, and concluded that the performance could be improved when films were prepared on substrates with special treatment. Previous studies mainly utilized glass or silicon plate as the substrate for CsI(Tl) scintillation screens. However, commercially available CsI(Tl) screens [8] mainly choose graphite, aluminum plate and fiber optic plate (FOP) as their scintillation screen substrates, without much explanation, and few reports focus on the role of substrates in the screen system.

In this paper, four substrates including $SiO_2$ plate ($SiO_2$-Subs.), silver-film coated $SiO_2$ plate (Ag-$SiO_2$-Subs.), graphite plate (C-Subs.) and fiber optic plate (FOP-Subs.) are chosen to fabricate CsI(Tl) films by thermal evaporation, and their imaging performance is evaluated by IEC standard 62220-1 [9]. C-Subs. has low density and strongly absorbs visible light photons, and thus, is used to absorb backward scintillation photons. On the other hand, $SiO_2$-Subs. is almost transparent in the visible region while silver film can reflect visible light, so both $SiO_2$-Subs. and Ag-$SiO_2$-Subs. are compared to verify the reflectivity of the silver film. In addition, FOP, which channels the scintillation photons, is also used to prepare CsI(Tl) screens.

**2. Experimental method**

A graphite disk was purchased from the Institute of Metal Research, Chinese Academy of Sciences, and a fused silica plate from Zhongcheng Quartz Co., Ltd. A fiber optic plate (FOP) was provided by Shanxi Changcheng Microlight Equipment Co., Ltd. The thickness of the selected plates was all 2 mm and the silver film was about 150 nm thick. The CsI(Tl) scintillation screens were fabricated by the thermal deposition method with the same evaporation parameters, so their morphologies were almost the same. The micro-column structure was confirmed by scanning electron microscope (SEM, HITACHI S4800), as shown in Fig. 1. For each of the substrates, three thicknesses of CsI(Tl) films were prepared: about 70 μm, 150 μm and 230 μm.

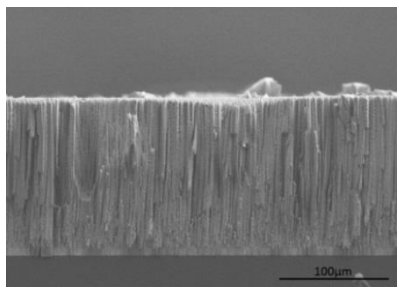

Fig. 1. Cross section of prepared CsI(Tl) scintillation film

Based on IEC 62220-1 standard, the measuring system was setup as described in Ref. [10]. The coupling structure is dependent on the substrates, as shown in Fig. 2. The values of X-ray tube voltage and additional filtration were set at RQA5, chosen from the radiation quality series as listed in Table 1. The exposure dose was set at 44.06 mGy, measured by a plane-parallel ionization chamber (UNIDOS E, PTW).

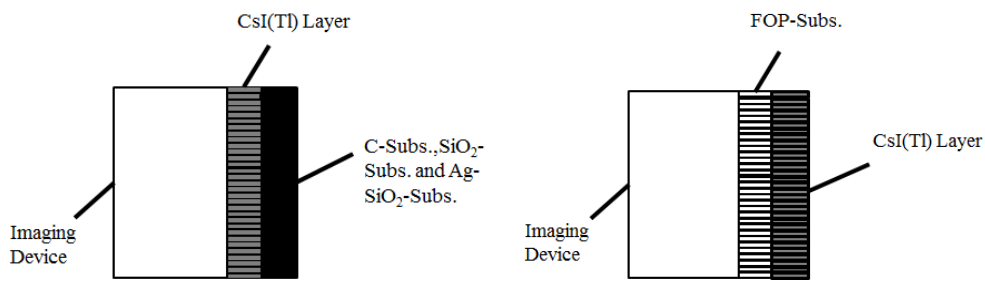

Fig. 2. CsI(Tl) screen coupling schematic for various substrates in X-ray imaging application.

As results, flat-field images and edge images are obtained for relevant image quality analysis. The evaluation characteristics are relative light output (RLO), modulation transfer function (MTF), normalized noise power spectrum (NNPS) and noise equivalent quanta (NEQ).

Table 1. Radiation quality specified in IEC 62220-1 standard [9].

| Radiation Quality No. | Approximate X-ray Tube Voltage (kV) | Half-Value Layer (HVL) (mm Al) | Additional Filtration (mm Al) | $SNR_{in}^2$ (Quanta per area per air kerma) [$1/(mm^2 \cdot \mu Gy)$] |
|---|---|---|---|---|
| RQA 3 | 50 | 4.0 | 10.0 | 21759 |
| RQA 5 | 70 | 7.1 | 21.0 | 30174 |
| RQA 7 | 90 | 9.1 | 30.0 | 32362 |
| RQA 9 | 120 | 11.5 | 40.0 | 31077 |

## 3. Results and Discussion

### 3.1 RLO

In order to compare the light output for various CsI(Tl) screens on different substrates, relative light output (RLO) is calculated using the following equation

$$\text{RLO} = \frac{LO_{sub,thick}}{LO_{C,70}}, \tag{1}$$

where LO $_{sub,\ thick}$, the average pixel value of the obtained flat-field images, stands for the light output of a scintillation layer of a given thickness on a certain substrate. For example, $LO_{C,\ 70}$ is the light output of a scintillation layer of 70 μm thickness on C-Subs..

As listed in Table 2, the RLO of CsI(Tl) film on C-Subs. is the lowest compared with the films on the other three substrates, because C-Subs. absorbs the backward scintillation photons the most among the selected substrates. However, the RLO of CsI(Tl) film on Ag-SiO$_2$-Subs. is larger than that of CsI(Tl) film on SiO$_2$-Subs. This change may be attributed to the pre-deposited Ag layer enhancing the reflection effect and reducing the number of scintillation photons penetrating the SiO$_2$-Subs. Since FOP-Subs. guides the scintillation photons, scintillation film on FOP-Subs. almost always yields the brightest image.

Table 2. RLO of CsI(Tl) films on various substrates and with different thicknesses

| RLO | Ag-SiO$_2$-Subs. | SiO$_2$-Subs. | C-Subs. | FOP-Subs. |
| --- | --- | --- | --- | --- |
| 70 μm | 3.8 | 3.1 | 1 | 4.6 |
| 150 μm | 7.1 | 6.0 | 2.8 | 7.4 |
| 230 μm | 10.6 | 7.1 | 4.3 | 8.7 |

**3.2 MTF**

Modulation transfer function (MTF) is a measure of the ability of an imaging detector to reproduce image contrast from subject contrast at various spatial frequencies [11-13]. Extensive experimental methods have been developed for the assessment of MTF of digital radiographic systems. In this study, MTF is measured by the slanted-edge method based on the IEC 62220-1 standard. One tungsten plate with thickness 1 mm and area 100 mm × 75 mm is utilized to obtain the edge image, from which the oversampled line spread function (LSF) is calculated. Finally the MTF is given by the following equation

$$\text{MTF}(f) = \frac{\text{FFT(LSF)}}{\text{FFT(LSF)}_{u=0}}, \tag{2}$$

where FFT is the digital fast Fourier transform and $f$ is the spatial frequency.

To study the substrate influence on image spatial resolution, the MTF curves of CsI(Tl) films on selected substrates with the same thickness (70 μm) under RQA5 radiation quality are shown in Fig. 4(a). It is found that the MTF curve of CsI(Tl) film on C-Subs. is the highest, while MTF curves on the other three substrates are almost the same. The probable cause is that C-Subs. suppresses scintillation photon reflection near the interface between the C-Subs. and the CsI(Tl) layer, so reduces the proportion of reflective photons, which travel a longer transmission distance, and thus improves the spatial resolution.

Additionally, the MTF curves of CsI(Tl) screen on C-Subs. as a function of thickness of CsI(Tl) layer are presented in Fig. 4(b). It is observed that as the thickness of scintillation layer increases, the MTF curves decrease accordingly. This correlation is in line with previous results reported by Zhao et al [10]. The probable reason is that thicker film means longer transmission distance, and induces greater lateral spreading of scintillation photons, which inevitably leads to worse spatial resolution.

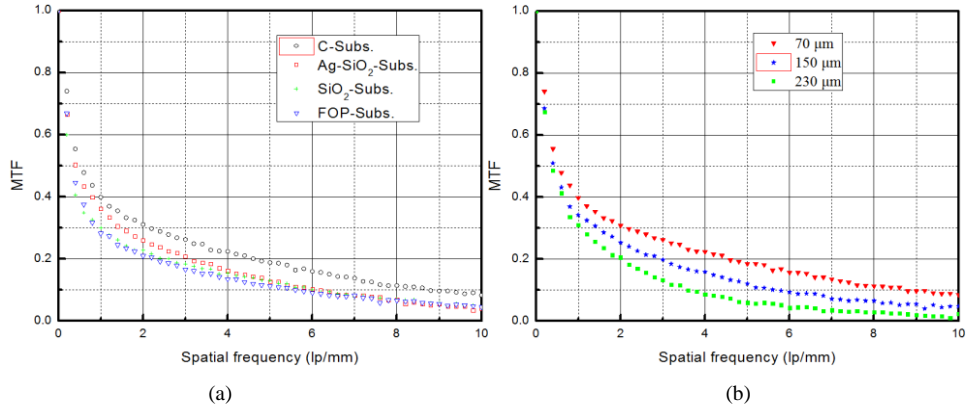

Fig. 3. (Color online) MTF curves of CsI(Tl) films on (a) various substrates, with 70 μm thickness; (b) C-Subs. with various thicknesses of CsI(Tl) layer

### 3.3 NNPS

The noise power spectrum (NPS) is defined as the modulus of the Fourier transform of the noise auto-covariance function [14]. It is often used to characterize the noise level of images obtained from X-ray imaging devices. The two dimensional (2D) NPS is calculated by Equation (3):

$$\text{NPS}(f, v) = \frac{\Delta x \cdot \Delta y}{M \cdot 256 \cdot 256} \sum_{m=1}^{M} |\text{FFT2}(I(x,y) - S(x,y))|^2, \quad (3)$$

where $\Delta x$, $\Delta y$ stand for the pixel spacing in the horizontal and vertical direction respectively, FFT2 is the 2D digital fast Fourier transform, M is the number of ROIs, I(x, y) is the pixel value and S(x, y) is the optionally fitted two-dimensional polynomial. For each flat-field image, a central region of 2048 × 2048 pixels is selected as the studied area. After converting 2D NPS to 1D NPS [15], the NNPS is

given by Equation (4):

$$\text{NNPS}(f) = \frac{\text{NPS}(f)}{\text{MVP}^2}, \tag{4}$$

where MVP is the mean value of the pixels used in the calculation of NPS. It is known that due to growth instability, the diameters of the CsI(Tl) needle-like crystal columns will inevitably be non-uniform, which will result in uneven light output between adjacent crystal columns and therefore a higher NNPS curve.

The NNPS curves of the CsI(Tl) scintillation screens as a function of spatial frequency with various thicknesses of scintillation film are compared in Fig. 5 (a, b, c). It is obvious that the CsI(Tl) film on FOP-Subs. demonstrates the best NNPS performance for all the thicknesses tested. The NNPS curves of scintillation films on SiO$_2$-Subs. and Ag-SiO$_2$-Subs., on the other hand, increase with the film thickness. The NNPS curves of scintillation film on C-Subs. are the worst among the selected substrates with film thickness of 70 μm and 230 μm, but the NNPS curve of 150 μm film on C-Subs. is significantly improved and better than that of SiO$_2$-Subs.

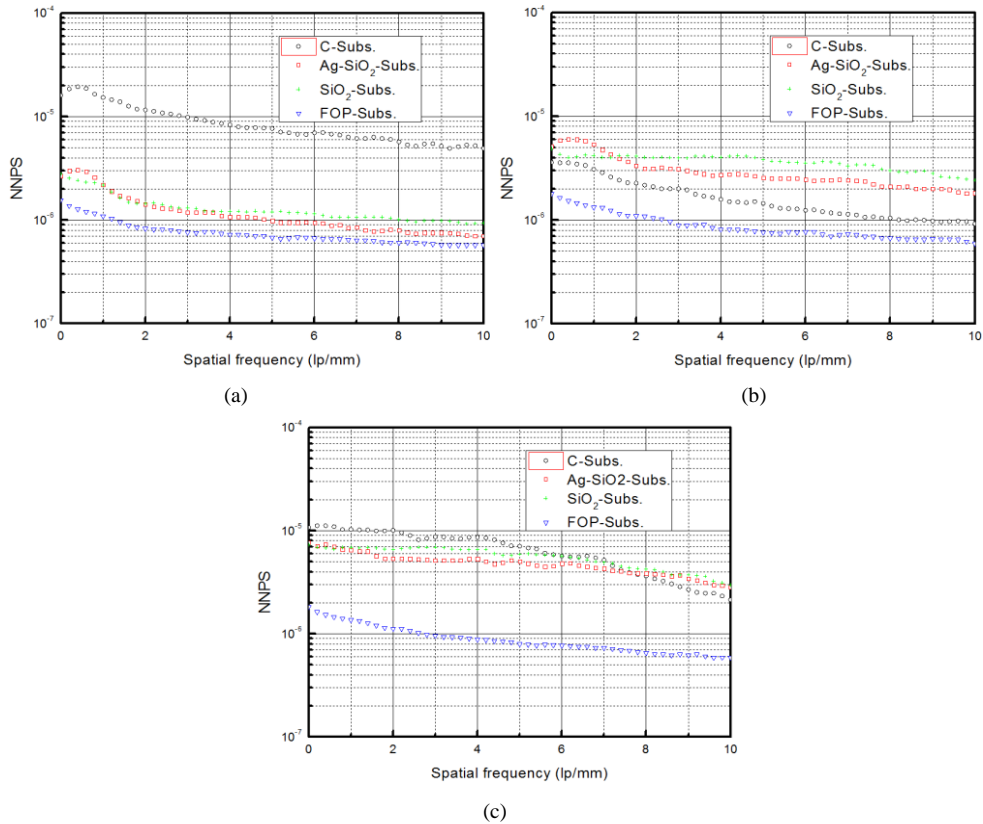

Fig. 4. (Color online) NNPS curves of CsI(Tl) films on selected substrates with thickness of (a) 70 μm, (b) 150 μm, (c) 230 μm

Compared with scintillation screens on the other three substrates, CsI(Tl) screen on the FOP-Subs. having the best NNPS may result from the tight coupling of the crystal columns with the fiber channels, guiding the scintillation photons uniformly. As for the remarkably high NNPS of 70 μm thick film on C-Subs., one possible reason is that its scintillation output is so low under the RQA5 exposure setup that the noise of the subsequent detector system becomes the main noise source and raises the noise level.

**3.4 NEQ**

Noise equivalent quanta (NEQ) is defined as the effective number of photons per cm of the detector [14]. The NEQ can be calculated by the following equation:

$$NEQ(f) = G^2 \frac{MTF(f)^2}{NPS(f)}, \tag{5}$$

where $f$ is the spatial frequency, $G$ is the system gain, MTF ($f$) is the modulation transfer function and NPS ($f$) is the noise power spectrum at RQA5 exposure level. The NEQ combines MTF with NNPS, which makes it a comprehensive index to assess the signal-to-noise ratio characteristic of a CsI(Tl) scintillation screen.

As shown in Figure 6, the NEQ curves of CsI(Tl) films on selected substrates are compared to investigate their imaging performance. Although the MTF of CsI(Tl) film on FOP-Subs. is relatively low, the NEQ curves remain steadily high in the various film thicknesses due to the low noise level. The scintillation film on C-Subs. yields relative high NEQ values as well, because its high NNPS curves are well compensated by it having the highest MTF level. However, it is obvious that the NEQ curves of scintillation films on $SiO_2$ plate with and without silver film are inferior to those with FOP-Subs. and C-Subs., due to their lower spatial resolution (MTF) and higher noise level (NNPS).

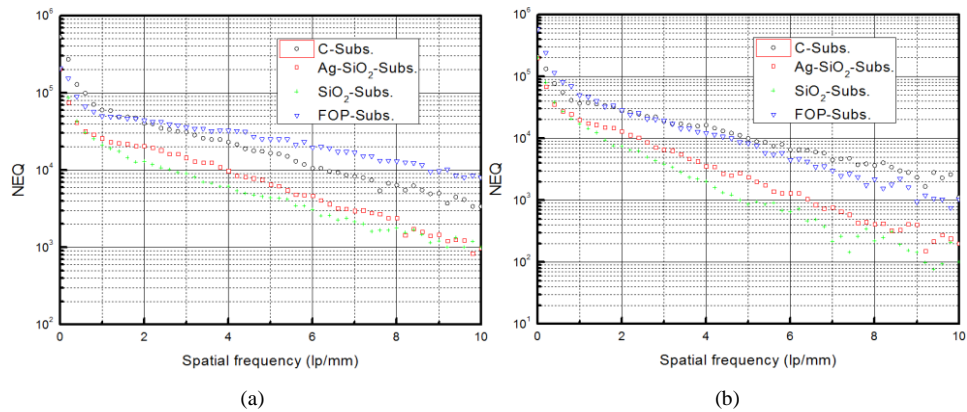

(a)      (b)

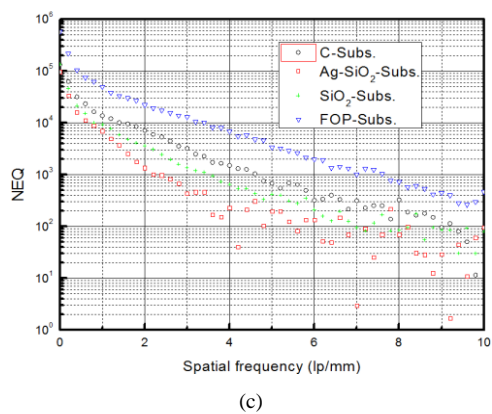

(c)

Fig. 5. (Color online) NEQ curves of CsI(Tl) films on chosen substrates with thicknesses of (a) 70 μm, (b) 150 μm, (c) 230 μm

## 4. Conclusion

CsI(Tl) scintillation screens based on four substrates were fabricated by the thermal deposition method and their imaging performance assessed by the IEC 62220-1 standard. It is indicated that the C-Subs. absorbs backward scintillation photons, resulting in the darkest image but with relatively high spatial resolution (MTF) and signal-to-noise ratio (NEQ). The silver film reflects the backward scattered photons and helps to increase the light output of the scintillation screen. However, the scintillation films based on $SiO_2$-Subs. have relatively low spatial resolution and signal-to-noise ratio, whether or not coated with silver film. Scintillation screens on FOP-Subs., however, due to the unique optical features of the fiber, can obtain bright images with low NNPS and high NEQ, with the only disadvantage being bad spatial resolution.